\documentclass[twocolumn,floatfix,superscriptaddress,prl, noeprint]{revtex4-2}

\usepackage{dsfont} % Typesetting the mathematical symbols for the natural numbers (N), whole numbers (Z), rational numbers (Q), real numbers (R) and complex numbers (C).
\usepackage{bm} % \bm to makfe bold fonts in math mode
\usepackage{mathtools,leftindex,tensor,mhchem}

\usepackage{bbold} %For identity
\usepackage{xcolor}
\usepackage{minitoc}
\usepackage{float}
\usepackage{physics}
\usepackage{graphicx,braket}
\usepackage{hyperref}
\hypersetup{colorlinks=true,linkcolor=blue,citecolor=blue}
\usepackage{amsmath,amssymb,amsfonts}
\usepackage{wrapfig}
\usepackage[export]{adjustbox}
\usepackage{cleveref} % To erence mul000tiple equations
\usepackage{hhline}
\usepackage{siunitx}
\usepackage{booktabs}% http://ctan.org/pkg/booktabs
\usepackage{enumitem}   

\usepackage{cleveref}% load last
\crefname{equation}{Eqs.}{Eqs.}
\Crefname{equation}{Equation}{Equations}% For beginning \Cref
\crefrangelabelformat{equation}{(#3#1#4--#5#2#6)}
\crefmultiformat{equation}{Eqs. (#2#1#3}{, #2#1#3)}{#2#1#3}{#2#1#3}
\Crefmultiformat{equation}{Equations (#2#1#3}{, #2#1#3)}{#2#1#3}{#2#1#3}

\begin{document}
	
	\allowdisplaybreaks % I write this to avoid huge gaps between equations and surrounding text, see https://tex.stackexchange.com/questions/21687/avoid-large-spaces-between-text-and-equations
	
	\flushbottom
	\title{Frequency-resolved Purcell effect for the dissipative generation of steady-state entanglement}

	\author{Alejandro Vivas-Via{\~n}a}
	\affiliation{Departamento de Física Teórica de la Materia Condensada and Condensed
		Matter Physics Center (IFIMAC), Universidad Autónoma de Madrid, 28049 Madrid,
		Spain}	
	\author{Diego Martín-Cano}
	\affiliation{Departamento de Física Teórica de la Materia Condensada and Condensed
		Matter Physics Center (IFIMAC), Universidad Autónoma de Madrid, 28049 Madrid,
		Spain}
	\author{Carlos S\'anchez Mu\~noz}
	\email[]{carlossmwolff@gmail.com}
	\affiliation{Departamento de Física Teórica de la Materia Condensada and Condensed
		Matter Physics Center (IFIMAC), Universidad Autónoma de Madrid, 28049 Madrid,
		Spain}
	
	\newcommand{\down}{\op{g}{e}}
	\newcommand{\up}{\op{e}{g}}
	\newcommand{\downd}{\op{+}{-}} % Down for the Dressed basis
	\newcommand{\upd}{\op{+}{-}}
	\newcommand{\app}{a^\dagger}
	\newcommand{\ssp}{\sigma^\dagger}
	\newcommand*{\Resize}[2]{\resizebox{#1}{!}{$#2$}}%
	\newcommand{\conc}{\mathcal C(\rho)}
%%%%%%%%%%%%%%%%%%%%%%%%% ABSTRACT %%%%%%%%%%%%%%%%%%%%%%%%%

\begin{abstract}
We report a driven-dissipative mechanism to generate stationary entangled $W$ states among strongly-interacting quantum emitters placed within a cavity. Driving the ensemble into the highest energy state---whether coherently or incoherently---enables a subsequent cavity-enhanced decay into an entangled steady state consisting of a single de-excitation shared coherently among all emitters, i.e., a $W$ state, well known for its robustness against qubit loss. The non-harmonic energy structure of the interacting ensemble allows this transition to be resonantly selected by the cavity, while quenching subsequent off-resonant decays. Evidence of this purely dissipative mechanism should be observable in state-of-the-art cavity QED systems in the solid-state, enabling new prospects for the scalable stabilization of quantum states in dissipative quantum platforms.
\end{abstract}
\date{\today} \maketitle

%%%%%%%%%%%%%%%%%%%%%%%%% INTRODUCTION %%%%%%%%%%%%%%%%%%%%%%%%%
Light-matter interfaces consisting of quantum emitters in solid-state cavity QED setups stand as promising platforms for the implementation of quantum technologies~\cite{KimbleQuantumInternet2008,OBrienPhotonicQuantum2009,LodahlInterfacingSingle2015,ToninelliSingleOrganic2021}. Examples of quantum emitters with optical transitions in the solid state include semiconductor quantum dots (QDs)~\cite{LodahlInterfacingSingle2015}, single molecules~\cite{ToninelliSingleOrganic2021}, or  color centres~\cite{SipahigilIntegratedDiamond2016,AwschalomQuantumTechnologies2018}. A key advantage of these systems is that the emitters can be integrated at fixed positions into photonic chips~\cite{OBrienPhotonicQuantum2009}, which enhances their prospects for scalability. It has been demonstrated that, by integrating these emitters in tailored photonic environments, such as photonic or plasmonic cavities and waveguides~\cite{PoyatosQuantumReservoir1996,VerstraeteQuantumComputation2009,LiuComparingCombining2016,ChangColloquiumQuantum2018}, the resulting collective dissipation can be used to stabilize entangled states of ensembles of emitters~\cite{PlenioCavitylossinducedGeneration1999,
	Gonzalez-TudelaEntanglementTwo2011, Martin-CanoDissipationdrivenGeneration2011, Gonzalez-TudelaMesoscopicEntanglement2013,RamosQuantumSpin2014,PichlerQuantumOptics2015,HaakhSqueezedLight2015,ChangColloquiumQuantum2018,ReitzCooperativeQuantum2022,kastoryano2011}, which represents a key hallmark for the development of quantum technologies, with key applications in quantum information~\cite{NielsenQuantumComputation2012,KimbleQuantumInternet2008,NarlaRobustConcurrent2016}  and metrology~\cite{BraskImprovedQuantum2015,PezzeQuantumMetrology2018}. 
However, the generation of entangled states in these platforms faces important challenges, such as the uncontrolled location of the emitters and the inhomogeneous broadening of their transition frequencies~\cite{GrimScalableOperando2019}, which prevents the emergence of collective phenomena either by direct dipole-dipole coupling or mediated by the photonic structures.  

Despite these challenges, recent years have witnessed a considerable progress in the fabrication of solid-state systems featuring collective quantum phenomena. Advances in the nanofabrication of photonic environments have enabled the observation of collective super- and subradiant dynamics between distant QDs~\cite{TiranovCollectiveSuper2023} and among molecules~\cite{rattenbacher2023}, as well as photon-mediated coherent interactions between
color centres in a nanocavity~\cite{EvansPhotonmediatedInteractions2018}. 
The eigenstates of strongly interacting quantum emitters offer a direct way to realize entanglement. In this direction, the small size of molecules and the possibility to synthesize bridged dimers with controllable intermolecular distances~\cite{DiehlEmergenceCoherence2014} allows to tackle the challenge of inherently weak dipole-dipole coupling by realizing strongly interacting emitters separated by sub-wavelength distances and featuring sub- and superradiant eigenstates~\cite{HettichNanometerResolution2002,TrebbiaTailoringSuperradiant2022,LangeSuperradiantSubradiant2023}.
 These advances are complemented by the development of scalable methods to address inohomogenous broadening by optically tuning different emitters into resonance~\cite{colautti2020}

The progress in synthesizing strongly interacting quantum emitters concurs with advancements in the design of light-matter interfaces with unprecedented radiative rate enhancements, which includes technologies such as plasmonic nanoantennas~\cite{ChikkaraddySinglemoleculeStrong2016,hoang2015,hoang2016} or hybrid cavity-antenna systems~\cite{GurlekManipulationQuenching2018,shlesinger2023}. This aligned development has called for further exploration into the phenomenology of strongly interacting quantum emitters when they are coupled to photonic structures~\cite{plankensteiner2017, plankensteiner2019}, with prospects such as enabling novel quantum-optical phases~\cite{parmee2020, perczel2017,ReitzCooperativeQuantum2022} or the fast and robust preparation of the emitter ensemble into its entangled eigenstates~\cite{NicolosiDissipationinducedStationary2004, Gonzalez-TudelaEntanglementTwo2011}.

In this Letter, we delve into this question and report  a driven-dissipative protocol to prepare steady entangled states in an ensemble of $N$ interacting quantum emitters coupled to a single mode cavity. 
This effect occurs when the coupling between emitters (in the absence of the cavity) yields hybridized eigenstates with a non-harmonic energy spectrum, such as the sub- and superradiant eigenstates for $N=2$. When this ensemble is pumped into its highest excited state ($|\frac{N}{2},\frac{N}{2}\rangle$ in the Dicke basis), the non-harmonic energy structure allows for a cavity to resonantly enhance a subsequent transition into an entangled $W$ state~\cite{dur2000} in the form of a symmetric superposition of all the one-de-excitation states, $|\frac{N}{2},\frac{N}{2}-1\rangle$.  
 Additional decay transitions, which fall out of resonance with the cavity, become suppressed, aiding in stabilizing the system into the desired $W$ state. This particular amplification of specific transitions within the ensemble is the reason behind terming it a frequency-resolved Purcell effect. The $W$ states prepared here have garnered particular interest and efforts for their generation~\cite{haffner2005,haas2014}, since they represent a fundamental class of entangled states~\cite{dur2000} which are persistent and robust against particle loss~\cite{briegel2001} and may display stronger non-classicality than GHZ states~\cite{sende2003}.

The mechanism that we propose works provided that the cavity linewidth is sufficiently narrow to resolve the energy structure of the interacting ensemble, while remaining in the bad-cavity limit with high cooperativity.
The process operates under continuous driving, which can be both coherent (for $N=2$)  and incoherent (for any $N$). The case of coherent driving of two emitters is extremely rich in phenomenology, giving rise to a variety of mechanisms of generation of entanglement which we discuss in more detail in the accompanying paper~\cite{Vivas-VianaDissipativeStabilization2023}.
\begin{figure}[t!]
	\includegraphics[width=0.45\textwidth]{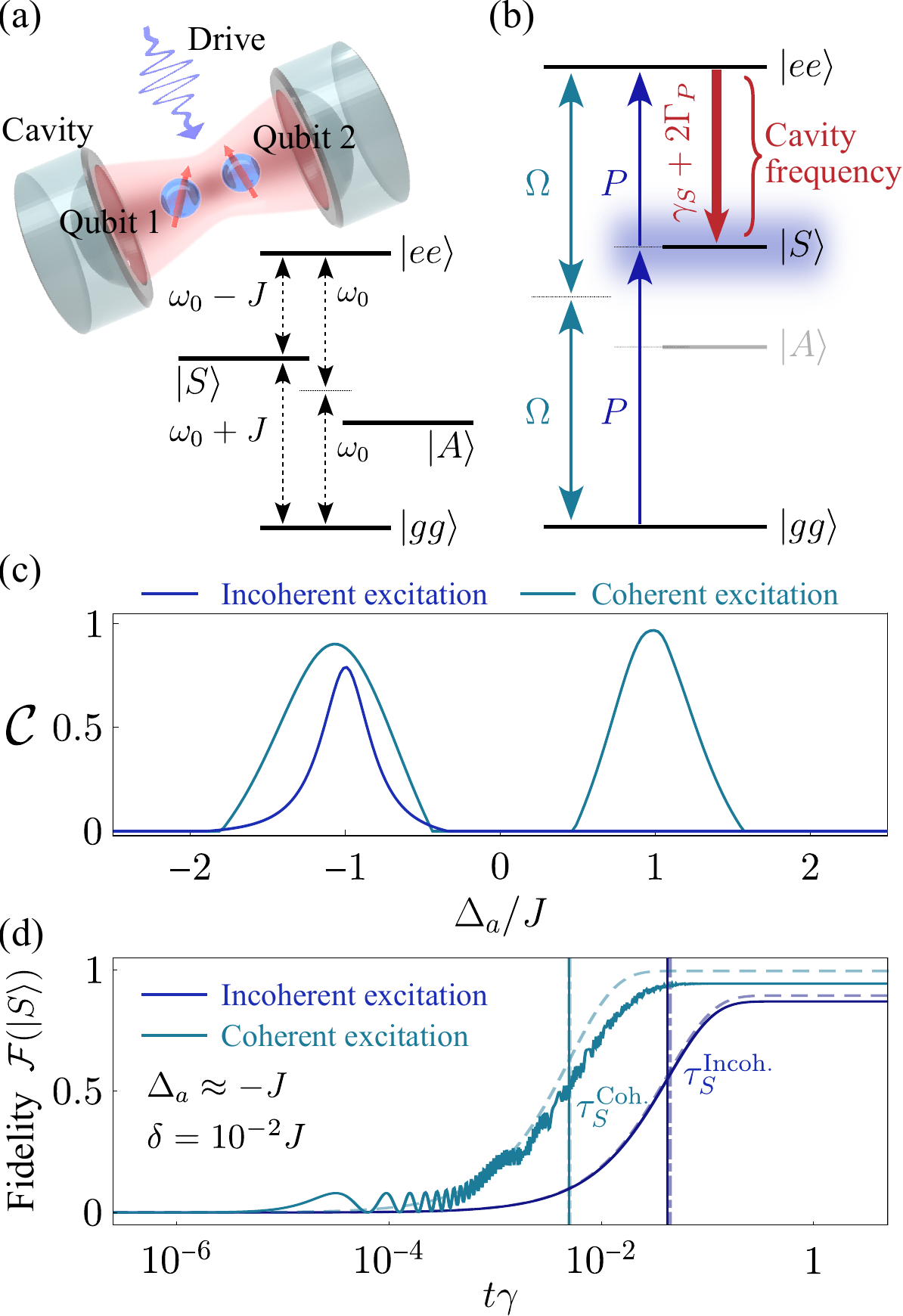}
	\caption{Entanglement generation for $N=2$ emitters.
 (a) Sketch of the system.
 (b) Energy diagram  and incoherent decay processes in the diagonal basis.
(c) Entanglement (measured by the concurrence) versus cavity detuning.
 (d) Time evolution of the fidelity $F(|S\rangle)$ when $\Delta_a=-J$. Dashed lines correspond to analytical predictions.
	Parameters: $\kappa=10^4\gamma$, $P=40\gamma$, $\Omega=10^4 \gamma$. }
	\label{fig:Fig1_Scheme}
\end{figure}

\textit{Model.---}
%%%%%%%%%%%%%%%%%%%%%%%%%
%
Each of the $N$ emitters is described as two-level system (TLS), i.e., a qubit, with annihilation operator $\hat\sigma_i$.
We start the paper by focusing on the case $N=2$, see Fig.~\ref{fig:Fig1_Scheme}(a).
For non-degenerate emitters, we express their natural frequencies as $\omega_1=\omega_0-\delta$ and $\omega_2=\omega_0+\delta$.
The single mode cavity is defined by its frequency $\omega_a$, annihilation operator $\hat a$, and coupling rate to the emitters $g$.
The coherent drive has a Rabi frequency $\Omega$ and it is resonant with the average qubit frequency, $\omega_L = \omega_0$. In the frame rotating at $\omega_0$ and under the rotating wave approximation, the system Hamiltonian reads (we set $\hbar=1$ henceforth) $\hat H= \hat H_q+ \hat  H_a+\hat  H_d$,
where $\hat H_q$ is the qubit-qubit Hamiltonian, 
$H_q= -\delta \hat \sigma_1^\dagger \hat \sigma_1+ \delta \hat \sigma_2^\dagger \hat \sigma_2+ J(\hat \sigma_1^\dagger \hat \sigma_2 + \text{H.c.} )$,
$\hat H_a$ is the cavity Hamiltonian, 
$\hat H_a=\Delta_a \hat a^\dagger \hat a+g [\hat a^\dagger (\hat \sigma_1+\hat \sigma_2)+\text{H.c.}]$,
and $\hat H_d$ is the coherent drive Hamiltonian, 
$\hat H_d= \Omega (\hat \sigma_1 +\hat  \sigma_2 + \text{H.c.})$,
where $\Delta_a\equiv \omega_a-\omega_0$ is the average qubit-cavity detuning, $J$ is the coherent coupling rate between the TLSs, and $\text{H.c.}$ denotes Hermitian conjugate.  

The dynamics of the density matrix is described, in the Markovian regime, by the master equation~\cite{GardinerQuantumNoise2004,FicekQuantumInterference2005,BreuerTheoryOpen2007},
$d\hat \rho/dt=-i [\hat H,\hat  \rho] 
+\frac{\kappa}{2}\mathcal{D}[\hat a]\hat \rho 
+ \sum_{i,j=1}^{N=2}\frac{\gamma_{ij}}{2}\mathcal{D}[\hat \sigma_i,\hat \sigma_j]\hat \rho +
\frac{P_i}{2}\mathcal{D}[\hat \sigma_i^\dagger ]\hat \rho $, 
where we have defined the Lindblad superoperators,
$
\mathcal{D}[\hat A, \hat B]\hat \rho \equiv 2\hat A \hat \rho \hat B^\dagger - \{\hat B^\dagger \hat A,\hat \rho \}
$
and
$
\mathcal{D}[\hat A]  \equiv \mathcal{D}[\hat A, \hat A].
$
Here, $\gamma_{ii} = \gamma$ and $P_i = P$ are the local rates of spontaneous emission and incoherent pumping, respectively, which we take equal for all the emitters; $\gamma_{12}=\gamma_{21}$ is the collective dissipative coupling rate between them;  and $\kappa$ is the photon leakage rate. 
The two pumping schemes considered in this work are set by fixing: (i) $\Omega= 0$, $P\neq 0$ (incoherent), and  (ii) $\Omega\neq 0$, $P=0$ (coherent), see Fig.~\ref{fig:Fig1_Scheme}(b). We note that the coherent drive excites directly the state $|ee\rangle$ via a two-photon resonance with an effective two-photon Rabi frequency, $\Omega_{\text{2p}}\equiv2\Omega^2 /J$~\cite{Vivas-VianaTwophotonResonance2021,Vivas-VianaDissipativeStabilization2023}.
We assume the bad-cavity limit~\cite{SavageStationaryTwolevel1988,CiracInteractionTwolevel1992,ZhouDynamicsDriven1998} ($\kappa \gg g \gg \gamma$), meaning that the qubit-cavity dynamics is irreversible and that the qubit decay is enhanced by the cavity, as quantified by the cooperativity $C\equiv \Gamma_P/\gamma = 4g^2/\kappa \gamma \gg 1$, where $\Gamma_P$ is the Purcell rate~\cite{PurcellResonanceAbsorption1946,KavokinMicrocavities2017}.

The mechanism that we report manifests in the regime of strongly interacting emitters $J\gg \delta$. For $N=2$, the basis of hybridized excitonic eigenstates is then the natural choice to describe the system, $\{ |gg\rangle, |A\rangle, |S\rangle, |ee\rangle  \}$, where $|S/A\rangle\equiv (|e g\rangle \pm |ge\rangle)/\sqrt{2}$ are entangled super- and subradiant eigenstates of the single-excitation subspace~\cite{FicekCooperativeEffects1986} [see Fig.~\ref{fig:Fig1_Scheme}(a)]. The corresponding eigenenergies and decay rates are then $E_\pm \approx \omega_0\pm J$ and $ \gamma_{S/A}\approx \gamma\pm \gamma_{12}$, respectively.
Unless stated otherwise, in all the results we set the following parameters: $\ J=9.18\times 10^4\gamma$, $\gamma_{12}=0.999\gamma,$  $\delta=10^{-2}J$ , $g=10^{-1}\kappa$. These values of $J$ and $\gamma_{12}$ correspond to what would be obtained from the dipole-dipole interaction between two emitters in a H-aggregate configuration in free space, separated by $r_{12}=2.5\ \text{nm}$ and with emission wavelength $ \lambda_0=780\, \text{nm}$~\cite{HettichNanometerResolution2002,TrebbiaTailoringSuperradiant2022,FicekQuantumInterference2005,CarmichaelStatisticalMethods1999,Vivas-VianaDissipativeStabilization2023}.
These high coupling values may only be applicable to a limited number of situations involving interacting molecules at nanometer distances~\cite{HettichNanometerResolution2002,DiehlEmergenceCoherence2014}. While we intentionally chose these values to clearly demonstrate the effect, it is important to note that entanglement can also be produced with less stringent parameters. We explore this aspect in more detail later in our work and in the Supplemental Material~\footnote{See Supplemental Material for further information.}.

\textit{Frequency-Resolved Purcell enhancement.---}
Once the stationary density matrix of the system ($\hat\rho_\text{ss}$) is obtained by solving the master equation, we quantify the degree of steady-state entanglement between the two qubits by means of the concurrence  $\mathcal{C}$~\cite{WoottersEntanglementFormation1998,WoottersEntanglementFormation2001,PlenioIntroductionEntanglement2007,HorodeckiQuantumEntanglement2009}. 
The mechanism to generate entanglement requires that the cavity enhances resonantly, via a frequency-resolved Purcell effect, the decay from the highest excited state to an entangled state where a single de-excitation is symmetrically distributed among all the emitters, which for $N=2$ is the superradiant state $|S\rangle$ [see Fig.~\ref{fig:Fig1_Scheme}(a-b)]. 
The process is enabled when the cavity frequency is set to $\Delta_a \approx -J$, matching the energy of the transition $|ee\rangle \rightarrow |S\rangle$. Figure~\ref{fig:Fig1_Scheme}(c) displays $\mathcal C$ versus the cavity detuning $\Delta_a$, confirming that, for both driving schemes, the system exhibits high values of concurrence at this resonance.

This mechanism requires a cavity with high cooperativity $C\gg 1$ and a linewidth small enough to resolve the different energy transitions taking place within the dimer, $ J \gg \kappa$. 
Indeed, under these conditions, and provided that the fine splitting due to two-photon dressing is not resolved by the cavity in the case of coherent driving, $\kappa \gg \Omega_\mathrm{2p}$~\cite{Vivas-VianaDissipativeStabilization2023}, 
the cavity can be adiabatically eliminated in both pumping schemes, yielding a simple effective Lindblad term that neatly describes the Purcell effect brought in by the cavity,
	$\mathcal L_\mathrm{cav}^\mathrm{eff} \approx \Gamma_P \mathcal{D}[\hat \xi_{S}]\hat \rho$,
with the effective jump operator $\hat \xi_{S}\equiv  |S\rangle \langle ee|$~\cite{Note1,Vivas-VianaDissipativeStabilization2023}. 
Crucially, activating the desired decay $|ee\rangle \rightarrow |S\rangle$ also requires an efficient excitation towards the state $|ee\rangle$, which we can achieve with both driving mechanisms, thanks to the two-photon resonant excitation in the coherent case, and to a sufficiently large $P\gg \gamma$ in the incoherent case.
The combination of this excitation and the subsequent cavity-enhanced decay gives rise to an effective incoherent pump of $|S\rangle$. 

\begin{figure}[t]
	\includegraphics[width=0.45\textwidth]{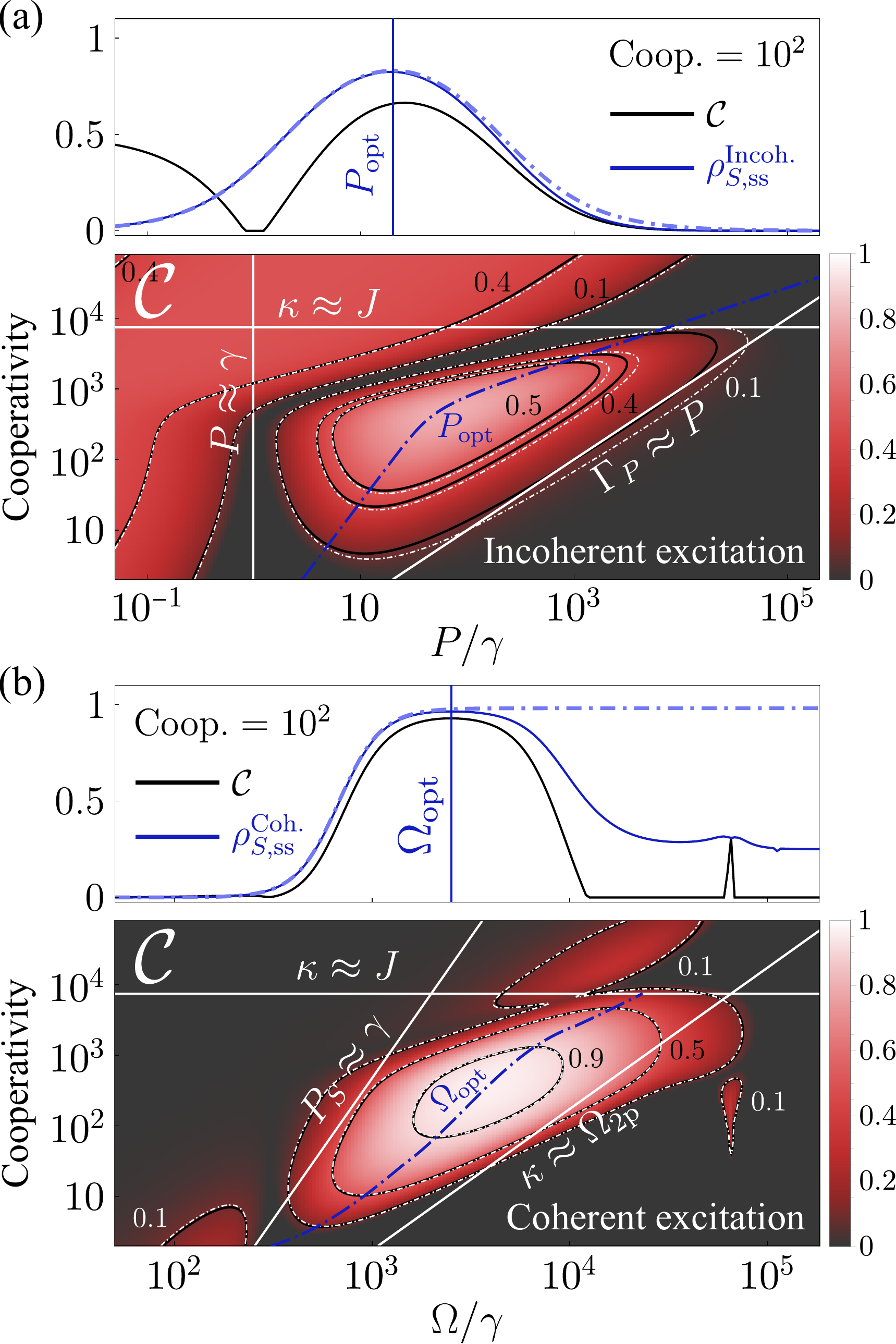}
	\caption{
 Stationary concurrence versus cooperativity $C$ and driving intensity: (a) incoherent excitation, (b) coherent excitation, for $N=2$ and  $\Delta_a\approx -J$. 
 White straight lines mark the conditions outlined in the text and bound the regions of formation of entanglement. Blue dot-dashed lines indicate the optimal driving strength that maximizes the concurrence.
 Black straight contours correspond to numerical calculations with a full model that perfectly match thee grey-dashed lines, obtained from an effective model where the cavity is adiabatically eliminated, see Ref.~\cite{Note1}.
 Top panels correspond to a cut of the concurrence (black) for a fixed cooperativity: $C=100$. Exact (blue) and analytical (dot-dashed blue) predictions of the population of $|S\rangle$ are depicted.
 }
	\label{fig:Fig3_Tunability}
\end{figure}

Such an incoherent mechanism may enable fast state preparations~\cite{hwang2009}, in contrast to the resonant driving of the transition $|gg\rangle \rightarrow |S\rangle$~\cite{TrebbiaTailoringSuperradiant2022}, that can never reach a stationary population inversion of the entangled state $|S\rangle$ and, furthermore, has a speed that is ultimately limited by the breakdown of the two-level system approximation through the excitation of higher electronic states.
The saturation of the stationary occupation probability to values close to unity will occur whenever the dissipative timescale of preparation, $\tau_S$, is much faster than the timescale of decay from $|S\rangle$ to other states, set by $\gamma^{-1}$. The resulting dynamics is shown in Fig.~\ref{fig:Fig1_Scheme}(d) in terms of the population of $|S\rangle$, which corresponds to the fidelity $\mathcal F(|S\rangle) \equiv \langle S|\hat\rho|S\rangle$~\cite{NielsenQuantumComputation2012}.

We obtained analytical estimations of $\tau_S$ by assuming the effective cavity-enabled decay given by $\hat\xi_S$. These are given by $
1/\tau_S^{\text{Incoh.}}\approx P+\Gamma_{P}-\sqrt{\Gamma_{P}^2+P\gamma_S},
$
and 
$
1/\tau_S^{\text{Coh.}}\approx (\Gamma_P-\text{Re}\sqrt{\Gamma_P^2-4\Omega_{\text{2p}}^2})/2
$,
for incoherent and coherent excitation respectively~\cite{Note1,Vivas-VianaDissipativeStabilization2023}. Figure~\ref{fig:Fig1_Scheme}(d) confirms the good agreement with exact numerical simulations.
The mechanism is most efficient when $\Gamma_P$ is the dominant dissipative rate; in that case, both timescales reduce to ${1/\tau_S^{\text{Incoh.}}\approx P}$ and ${1/\tau_S^{\text{Coh.}}\approx P_\mathrm{S}}$, where $P_S \equiv \Omega_\mathrm{2p}^2/\Gamma_P$ can be considered an effective incoherent pumping rate of the state $|S\rangle$ in the case of coherent excitation~\cite{Vivas-VianaDissipativeStabilization2023}. The ability to make $\tau_S$ small by designing systems with large Purcell factor and fixing high pumping rates has important implications for the generation of entanglement in the presence of extra decoherence channels, as we consider in detail in Ref.~\cite{Note1}. We find that entanglement can be consistently generated provided $\tau_S\ll \gamma_\text{d}^{-1}$, where $\gamma_\text{d}$ is the relevant decoherence rate. This condition also ensures that entanglement survives as a steady state, with the exception of the particular case of local pure dephasing.

Our understanding of the mechanism suggests a hierarchy of rates that should be respected for it to occur. A general condition is $J\gg \kappa \gg \Gamma_P \gg \gamma$, which ensures a frequency-dependent Purcell effect ($J\gg \kappa \gg \Gamma_P$) with high cooperativity ($ \Gamma_P \gg \gamma$). Furthermore, there are conditions that depend on the driving mechanism and that ensure that the target entangled state is efficiently populated for 
(i) 
incoherent excitation: $\Gamma_P \gg P \gg \gamma $, and (ii)
  coherent excitation: $P_S \gg \gamma$ and $\kappa \gg \Omega_\mathrm{2p}$.

Figure~\ref{fig:Fig1_Scheme}(c) also shows that, in the coherent excitation scheme, the system features an antisymmetric entanglement resonance $\Delta_a\approx J$ due to the population of the subradiant eigenstate $|A\rangle$ when $\delta$ is small but finite. We discuss this process in more detail in Ref.~\cite{Vivas-VianaDissipativeStabilization2023}.
This mechanism is highly inneficient for incoherent excitation due the depletion of the antisymmetric state by the incoherent repumping.

\textit{Dependence with pumping strength.---}
The mechanism we present is optically tunable via the intensity of the pump. To explore this dependence on a range of platforms, we compute the stationary concurrence in terms of the driving intensity ($P$ or $\Omega$) and the cooperativity ($C$) see Fig.~\ref{fig:Fig3_Tunability}. 
The regions of high entanglement  are enclosed within the white lines that delineate the set of conditions outlined previously, confirming their validity. Notably, we find high values of the concurrence in the range of cooperativities $C\sim 10-10^2$, accessible by state-of-the-art open-cavity configurations used, e.g., in reports of strong coupling with single molecules~\cite{pscherer2021}.
A key parameter is the optimal pumping rate that maximizes entanglement for a given cavity setup. For the incoherent pumping case, we find that this is given by 
$
P_\text{opt}=(\Gamma_P/2)\sqrt{(\kappa/J)^2+ 16/C}
$ (see Ref.~\cite{Note1} for details), whose accuracy is confirmed in Fig.~\ref{fig:Fig3_Tunability}(a).
The numerically determined optimum value for the coherent drive, $\Omega_{\text{opt}}$, is shown in Fig.~\ref{fig:Fig3_Tunability}(b).

\begin{figure}[t]
	\includegraphics[width=0.47\textwidth]{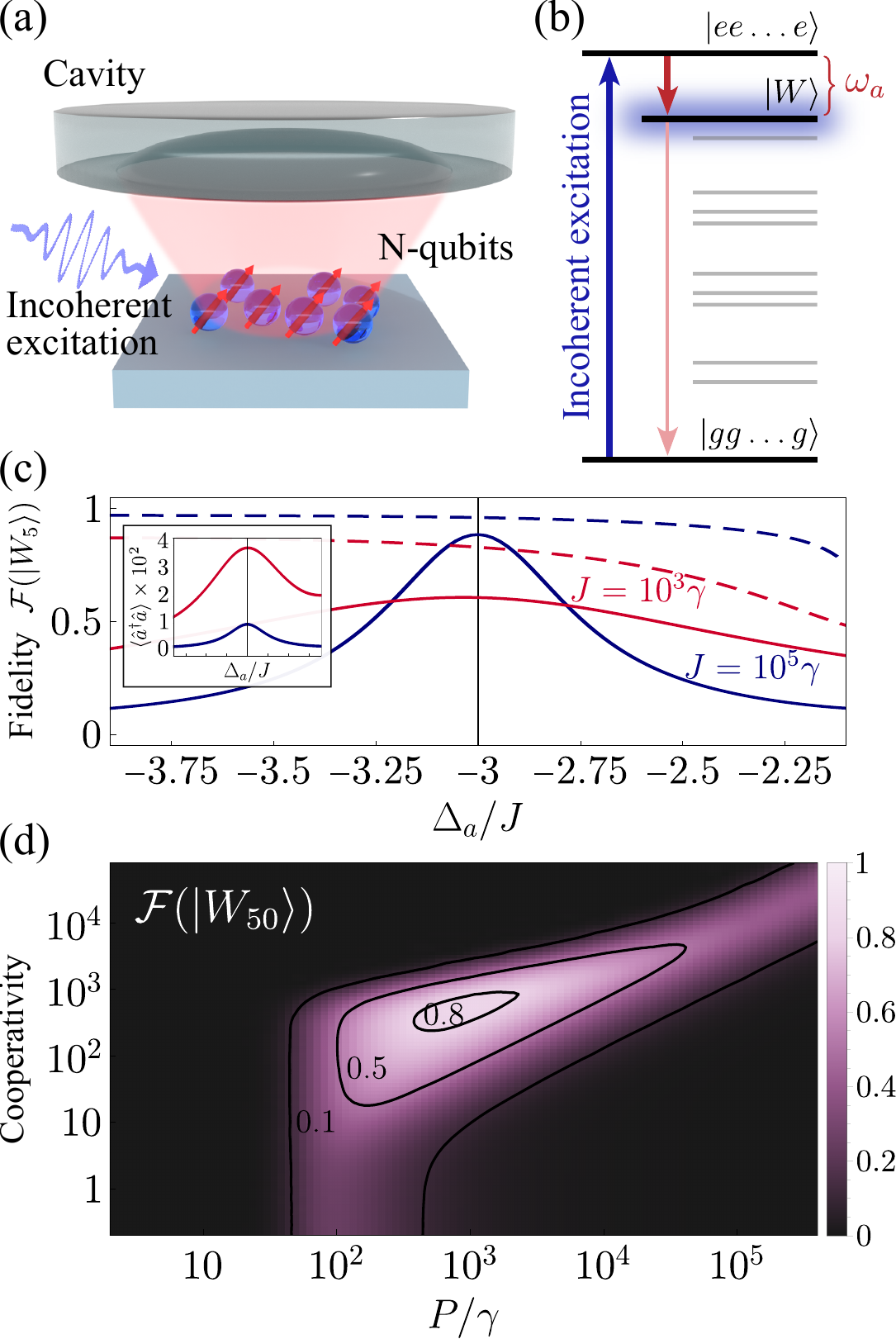}
	\caption{(a) Sketch of the setup of entanglement generation for $N>2$. (b) Enhanced dissipative transition in the Dicke ladder.
 (c) Fidelity with the $W$ state versus cavity detuning for $N=5$ for two values of $J$. The corresponding cavity population is shown in the inset. Dashed lines show the fidelities for a post-selected state, heralded by the detection of a photon from the cavity.
 (d) Fidelity with the $W$ state for $N=50$ versus cavity cooperativity and incoherent pumping rate.
	}
	\label{fig:Fig4_Generalization}
\end{figure}

\textit{Generalization to $N$ emitters.---} Figure~\ref{fig:Fig3_Tunability} suggests that, for a fixed set of cavity parameters, coherent driving yields higher values of entanglement than the incoherent case. However, incoherent driving allows to straightforwardly extend the scheme to the case of $N>2$  [Fig.~\ref{fig:Fig4_Generalization}(a)]. As an illustrative example, we consider $N$ degenerate emitters with an all-to-all interaction Hamiltonian, as could be provided by the coupling to a common photonic environment~\cite{goldstein1997,EvansPhotonmediatedInteractions2018,ChangColloquiumQuantum2018}. In the frame rotating at their bare frequency, the Hamiltonian reads $\hat H_q = J \hat S^+\hat S^- $, with $\hat S^-\equiv \sum_i^N \hat\sigma_i $. In addition, we assume a coupling to the cavity $g \hat a^\dagger \hat S^- + \text{H.c.}$, and the same type of dissipators considered before. The eigenstates of $\hat H_q$ are the Dicke states $|j,m\rangle$~\cite{DickeCoherenceSpontaneous1954,ShammahOpenQuantum2018}, with eigenvalues $\lambda_{j,m} = J[j(j+1) - m(m-1)]$. 

As for $N=2$, qubit-qubit interactions provide a non-equidistant set of eigenvalues, enabling the tuning of the cavity to enhance specific transitions within the ladder, energetically separated from each other by values $\propto J$.
Following our previous discussion, a strong pump $P$  will tend to populate the highest excited state, $|N/2,N/2\rangle = |ee\ldots e\rangle$, allowing the cavity to successively enhance the transition from this state to a $W$ entangled state of the form $|W_N\rangle \equiv |N/2,N/2-1\rangle =(|ge\ldots e\rangle + |eg\ldots e\rangle)+\ldots + |ee\ldots g\rangle)/\sqrt N$ [see Fig.~\ref{fig:Fig4_Generalization}(b)]. This specific transition is selected by setting the cavity frequency at $\Delta_a = J(2-N)$. Figure~\ref{fig:Fig4_Generalization}(c) shows the fidelity between the steady-state and the $W$ state,    $\mathcal F(|W_N\rangle)$,  for $N=5$ versus cavity detuning, confirming that high fidelity values are achieved at resonance. Notably, the activation of this mechanism translates into measurable properties of the emission by the cavity. This is shown in the inset of Fig.~\ref{fig:Fig4_Generalization}, which depicts how the resonant profile of the fidelity correlates with the same profile in the cavity population, proportional to emission intensity. This change is accompanied by a change of the photon counting statistics, that becomes antibunched at resonance even when many emitters are involved (shown in Ref.~\cite{Note1}), suggesting that these features could work as evidence of the formation of entanglement.

The plot also illustrates how a decrease of two orders of magnitude in $J$ results in a moderate decrease in fidelity, from approximately $0.9$ to about $0.6$ (a more systematic analysis of how the efficiency of the mechanism depends on $J$ is provided in Ref.~\cite{Note1}).

Furthermore, we show evidence that this mechanism can be used in a projective preparation scheme based on post-selection~\cite{haas2014}. In particular, we show fidelities  for the conditional state heralded by the detection of a photon in the steady-state, $\hat\rho_c = \hat a\hat\rho_\text{ss}\hat a^\dagger/\text{Tr}[\hat a\rho_\text{ss}\hat a^\dagger]$, which always shows a significant increase in fidelity with respect to the non-post-selected case. Figure~\ref{fig:Fig4_Generalization}(d) demonstrates how high values of the fidelity---that follow the patterns previously discussed in Fig.~\ref{fig:Fig3_Tunability}(a)---can be achieved for a number of emitters as high as $N=50$ (simulations were done using the PIQS library in QuTiP~\cite{ShammahOpenQuantum2018,johansson2012,johansson2013}).

Our results also imply that strongly interacting molecules located at subwavelength distances, which have been shown to support entangled eigenstates~\cite{Hettich2002, TrebbiaTailoringSuperradiant2022, LangeSuperradiantSubradiant2023}, could serve as suitable systems to observe these effects when coupled to state-of-the-art Fabry-Perot open cavities~\cite{pscherer2021}. We anticipate that one of the primary challenges in observing the mechanism described here is the necessity for a substantial disparity between the coherent coupling ($J$) and spontaneous emission ($\gamma$) rates, which occupy opposite ends of the required rate hierarchy. Our results prompt further investigation into the exotic non-equilibrium states that can arise in novel light-matter interfaces where the coupling between quantum emitters and/or the supression of undesired emission plays a crucial role, such as emitter interactions within a photonic band gap~\cite{ArcariNearUnityCoupling2014, ChangColloquiumQuantum2018}, implementations of giant atoms~\cite{kockum2018,kannan2020} or subwavelength atomic arrays~\cite{parmee2020}.

%\acknowledgments
\textit{Acknowledgments.---}
We acknowledge financial
support from the Proyecto Sin\'ergico CAM 2020 Y2020/TCS-
6545 (NanoQuCo-CM), and MCINN projects PID2021-126964OB-I00 (QENIGMA) and TED2021-130552B-C21 (ADIQUNANO). C. S. M. and D. M. C. acknowledge 
the support of a fellowship from la Caixa Foundation (ID 100010434), from the European Union's Horizon 2020 Research and Innovation Programme under the Marie Sklodowska-Curie Grant Agreement No. 847648, with fellowship codes  LCF/BQ/PI20/11760026 and LCF/BQ/PI20/11760018. D. M. C. acknowledges support from the Ramon y Cajal program (RYC2020-029730-I).

\let\oldaddcontentsline\addcontentsline% Store \addcontentsline
\renewcommand{\addcontentsline}[3]{}% Make \addcontentsline a no-op
\bibliography{Refs_AVV, Entanglement-dimers}

\let\addcontentsline\oldaddcontentsline% Restore \addcontentsline

\widetext
\clearpage
%%%%%%%%%% Merge with supplemental materials %%%%%%%%%%
%%%%%%%%%% Prefix a "S" to all equations, figures, tables and reset the counter %%%%%%%%%%
\setcounter{equation}{0}
\setcounter{figure}{0}
\setcounter{table}{0}
% \makeatletter
\renewcommand{\theequation}{S\arabic{equation}}
\renewcommand{\thefigure}{S\arabic{figure}}
%\renewcommand{\bibnumfmt}[1]{[S#1]}
%\renewcommand{\citenumfont}[1]{S#1}

%%%%%% SUPPLEMENTAL MATERIAL %%%%%%

\begin{center}
	\textbf{\large Supplemental Material:}\\
	\vspace{0.1cm}
	\textbf{ \large Frequency-resolved Purcell effect for the dissipative generation of steady-state entanglement} \\
	\vspace{0.3cm}
	Alejandro Vivas-Viaña, Diego Martín-Cano, and Carlos Sánchez Muñoz.
	\vspace{0.2cm}
\end{center}

\setcounter{page}{1}
\setcounter{secnumdepth}{1}

\
\tableofcontents

\section{Adiabatic elimination of the cavity}
\label{sec:SM_Lindblad}

Here, we prove that the cavity contribution to the dynamics can be described by the effective Lindblad term, $\mathcal{L}_{\text{cav}}^{\text{eff}}\approx \Gamma_P \mathcal{D}[\hat \xi_S]\hat \rho$, in the case of two quantum emitters, $N=2$.
We focus on the case of a incoherent excitation, where the drive does not dress the energy level structure of the emitter ensemble. The case of coherent driving is discussed in more detail in Ref.~\cite{Vivas-VianaDissipativeStabilization2023}. 

Under a Born-Markov approximation, the effective contribution of the cavity is accurately described by the following Bloch-Redfield term~\cite{Vivas-VianaDissipativeStabilization2023},
\begin{equation}
    \mathcal{L}_{\text{cav}}^{\text{eff}}\equiv \sum_{i,j,m,n=1}^4 \left( \frac{g_{ij}g_{mn}^* }{\kappa/2+i(\Delta_a-\omega_{ij})}\left[\hat \sigma_{ij}\hat \rho,\hat \sigma_{mn}^\dagger\right] + \text{H.c.} \right),
    \label{eq:Nakajima}
\end{equation}
where we defined $\hat \sigma_{ij}=|j\rangle \langle i|$, $g_{ij}=g\langle j |\hat \sigma_1+ \hat \sigma_2 |i\rangle$, and $\omega_{ij}\equiv \lambda_i - \lambda_j$, where $|i\rangle$ and $\lambda_i$, $i=1,\ldots,4$, are the eigenvectors and eigenvalues of the qubit-qubit system.
In order to keep the discussion as general as possible, we consider in this Section  the possibility of a finite qubit-qubit detuning $\delta$, so that the single-excitation eigenstates are given by $|\pm \rangle \equiv 1/\sqrt{2}(\sqrt{1\mp \sin \beta} |eg\rangle\pm \sqrt{1\pm \sin \beta}|ge\rangle) $, where $\beta\equiv\arctan (\delta/J)$ denotes a mixing angle. The corresponding eigenvalues are  $\omega_\pm =\Delta\pm R$, where $R=\sqrt{J^2 +\delta^2}$ is the Rabi frequency of the emitter-emitter coupling. We assume in the follwing that $\delta$ is finite but much smaller than $J$, so that $R\approx J$.
The eigenstates of the qubit-qubit system are labeled as
\begin{equation}
    |1\rangle = |gg\rangle;\quad
    |2\rangle = |-\rangle;\quad
    |3\rangle = |+\rangle;\quad
    |4\rangle = |ee\rangle.
\end{equation}

By setting the cavity frequency at the symmetric resonance, the terms where $\omega_{ij}\approx \Delta_a \approx -J$ will be dominant in the sum over $(i,j)$ in Eq.\eqref{eq:Nakajima}.
These terms are given by the following set of enhanced transitions,
\begin{equation}
    \mu_S \equiv \{ (2,1), (4,3) \},
\end{equation}
corresponding to  $|-\rangle \rightarrow |gg\rangle$, and $|ee\rangle \rightarrow |+\rangle$, respectively. The rest of the terms in the sum can be neglected since they are proportional to $g^2/J\rightarrow 0$, provided the conditions for a frequency-resolved Purcell effect hold,
\begin{equation}
    J\gg \kappa \gg g.
    \label{eq:conditions_freq_Purcell}
\end{equation}
By performing a rotating-wave approximation, terms not belonging to $\mu_S$ in the sum over $(m,n)$
 can also be neglected, since these will rotate at frequencies $J\gg g^2/\kappa$ in the Heisenberg picture [the inequality being a consequence of Eq.~\eqref{eq:conditions_freq_Purcell}].  

Under these conditions, the cavity contribution in Eq.~\eqref{eq:Nakajima} can be reduced to the sum
\begin{equation}
    \mathcal{L}_{\text{cav}}^{\text{eff}}\approx  \frac{g_{21}}{\kappa/2}[\hat \sigma_{21} \hat \rho,\hat \Lambda]+ \frac{g_{43}}{\kappa/2}[\hat \sigma_{43}\hat \rho,\hat \Lambda]+ \text{H.c.},
\end{equation}
where 
\begin{equation}
   \hat  \Lambda\equiv \sum_{(m,n) \in \mu_S} g_{mn}^* \hat \sigma_{mn}^\dagger=g_{21}^* \hat \sigma_{21}^\dagger+g_{43}^* \hat \sigma_{43}^\dagger\approx \beta\frac{g}{\sqrt{2}} |-\rangle \langle gg | + \sqrt{2}g |ee\rangle \langle +|,
\end{equation}
where $g_{21}\approx \beta g/\sqrt{2}$ (assuming $\delta \ll J$), $g_{43}\equiv \sqrt{2}g$, $\hat \sigma_{21}\equiv |gg\rangle \langle - |$, and $\hat \sigma_{43}\equiv |+\rangle \langle ee|$. 
Rearranging these terms, we obtain an effective Lindblad-like term of the form,
\begin{equation}
	\mathcal{L}_{\text{cav}}^{\text{eff}} \approx \Gamma_P \mathcal{D}[\hat \xi_{S}]\hat \rho,
 \label{eq:liouvillian_cav}
\end{equation}
with $\Gamma_P \equiv 4g^2/\kappa$ and a single jump operator given by
\begin{equation}
    \hat \xi_S\equiv |+\rangle \langle ee|+ \beta |gg\rangle \langle -|.
    \label{eq:xi_S}
\end{equation}
Notice that, in the limit $\beta\rightarrow 0$ considered in the main text, we obtain $\hat\xi_S \approx |S\rangle \langle ee|$. The effective dynamics $\mathcal L^\text{eff}_\text{cav}$ thus induces an enhanced superradiant transition from the doubly-excited state to the symmetric state, $|ee\rangle \rightarrow |S\rangle$, with a rate $2\Gamma_P$. For non-neglibible $\delta$, the second term in Eq.~\eqref{eq:xi_S} corresponds to a weakly enhanced  transition from the subradiant state to the ground state, $|-\rangle \rightarrow |gg\rangle$, with a rate $\beta^2 \Gamma_P$.
Given its quadratic dependence with the small parameter $\beta$, this second type of transition will clearly be negligible in the dimer configuration $\delta \ll J$, where $\beta=\tan^{-1}(\delta/J)\ll 1$.

\section{Analytical solutions of the master equation under incoherent excitation}
\label{sec:SM_AnalyticalDensityMatrix}
In the case of an incoherent drive and $N=2$, we can obtain good analytical approximations of the dynamics based on the adiabatic elimination of the cavity discussed above. A first approach is the direct diagonalization of the Bloch-Redfield equation in Eq.~\eqref{eq:Nakajima}, which allows to obtain analytical solutions of the steady state of the emitters, provided $\kappa \gg g$. 
In the limit $\beta\approx0$ and setting $\Delta_a\approx -J$, the steady-state populations read (assuming $P\gtrsim \gamma$)
\begin{align}
\label{eq:rho_S_general}
\rho_{S,\text{ss}}&\approx \frac{P^2 J^2 (8g^2 + \gamma_S \kappa)}{8 g^2 P J^2(P+\gamma_S) + 4g^4 P \kappa +J^2\kappa P^3}, \\
\rho_{A,\text{ss}}&\approx \frac{P \left[ 4g^2J^2 \gamma_S+2g^4 \kappa+J^2 \kappa \gamma_S \gamma    \right]}{8 g^2 P J^2(P+\gamma_S) + 4g^4 P \kappa +J^2\kappa P^3}, \\
\rho_{ee,\text{ss}}&\approx \frac{P^2 J^2 \kappa (P+\gamma)}{8 g^2 P J^2(P+\gamma_S) + 4g^4 P \kappa +J^2\kappa P^3}.
\end{align}
We can rewrite Eq.~\eqref{eq:rho_S_general}, which corresponds to the relevant fidelity $\mathcal F(|S\rangle)$ used as a figure of merit in the text, in terms of the Purcell rate $\Gamma_P$. In the limit $C\gg 1$, it takes the following transparent form
\begin{equation}
\label{eq:rho_S_general_simplified}
    \rho_{S,\text{ss}} \approx \left[1 + \frac{P}{2\Gamma_P} + \frac{\gamma_S}{P} + \frac{1}{8}\frac{\Gamma_P}{P}\left(\frac{\kappa}{J} \right)^2 \right]^{-1}.
\end{equation}
This expression clearly shows how, under the conditions established in the main text for the efficient generation of entanglement, $J \gg \kappa \gg \Gamma_P \gg P \gg \gamma$, the population saturates to its maximum possible value $\rho_{S,\text{ss}}\approx 1$. 

The expressions above are general and apply beyond the conditions of the frequency-resolved Purcell effect in Eq.~\eqref{eq:conditions_freq_Purcell}. However, using the simple form of the cavity-induced Lindblad term in Eq.~\eqref{eq:liouvillian_cav}, which applies in the frequency-resolved Purcell regime, allows us to obtain simpler expressions for the steady state and the dynamics. In particular, in the dimer configuration $\beta\approx 0$, the system can be reduced to a cascaded three-level system composed of $\{ |gg\rangle, |S\rangle, |ee\rangle\}$, since the subradiant state, $|A\rangle$, can be neglected as it is essentially decoupled from the system dynamics due to its dark nature~\cite{FicekQuantumInterference2005}. Consequently, the system is effectively described by a density matrix $\hat \rho_{\text{eff}}$ whose dynamics is governed by the following master equation
\begin{equation}
    \frac{d \hat \rho_{\text{eff}}}{dt}=  \left(
    \frac{\gamma_S}{2} \mathcal{D}[\hat \sigma_{gg,S}] +
    \Gamma_P \mathcal{D}[\hat \sigma_{S,ee}] +
    \frac{P}{2} \mathcal{D}[\hat \sigma_{gg,S}^\dagger] +
    \frac{P}{2} \mathcal{D}[\hat \sigma_{S,ee}^\dagger]\right) \hat \rho_{\text{eff}}.
\end{equation}
where we have defined $\hat \sigma_{gg,S}\equiv |gg\rangle \langle S|$ and $\hat \sigma_{S,ee}\equiv |S\rangle \langle ee|$.
From this equation, we can straightforwardly compute analytical expressions for the qubit-qubit dynamics, such as the symmetric state population $\rho_{S}(t)\equiv\langle S | \hat \rho_{\text{eff}} (t) |S\rangle$,
\begin{equation}
    \rho_{S}(t)\approx \rho_{S,\text{ss}} \left( 1- e^{-t/\tau_S }\right),
\end{equation}
where $\rho_{S,\text{ss}}$ is its stationary population,
\begin{equation}
    \rho_{S,\text{ss}}\approx \left[1+ \frac{P}{2 \Gamma_{P}}+\frac{\gamma_S}{P}\right]^{-1},
    \label{eq:rhoS_purcell}
\end{equation}
and $\tau_S$ denotes the stabilization timescale,
\begin{equation}
    1/\tau_S^{\text{Incoh.}} \approx  P+\Gamma_P+\frac{\gamma_S}{2}-\sqrt{P\gamma_S+\frac{1}{4}(\gamma_S-2\Gamma_P)^2} \approx P+\Gamma_{P}-\sqrt{\Gamma_{P}^2+P\gamma_S}.
\end{equation}
The last approximation corresponds to the expression presented in the text under the assumption $\Gamma_P\gg \gamma_S$. We note  that Eq.~\eqref{eq:rho_S_general_simplified} corresponds to the limit $J \gg \kappa$ of Eq.~\eqref{eq:rhoS_purcell}.
For analytical expressions of the stationary density matrix under coherent excitation, we refer the reader to Ref.~\cite{Vivas-VianaDissipativeStabilization2023}.

These results allow us to obtain an analytical expression for the optimal incoherent pumping rate $P_{\text{opt}}$, that maximizes the population of the symmetric state $|S\rangle$. 
To do so, we find the $P$ that maximizes the expression for $\rho_{S,\text{ss}}$ in Eq.~\eqref{eq:rho_S_general}, which has a wider range of validity than Eq.~\eqref{eq:rhoS_purcell}. The optimal pumping rate thus reads, assuming $\gamma_S = 2\gamma$ and the cooperativity $C\equiv \Gamma_P/\gamma$,
\begin{equation}
	P_{\text{opt}}=\frac{\Gamma_P}{2}\sqrt{\left(\frac{\kappa}{J}\right)^2+\frac{16}{C}}.
\end{equation}
Under the conditions that we established for the efficient generation of entanglement, $J\gg \kappa$ and $C\gg 1$ (equivalent to $\Gamma_P \gg \gamma$), this equation  shows that the optimum pumping consistently fulfills the condition $P\ll \Gamma_P$. By substituting $P_\text{opt}$ back in Eq.~\eqref{eq:rho_S_general_simplified}, we can obtain $\rho_{S,\text{ss}}^\text{max}$, the maximum achievable population of $|S\rangle$:
\begin{equation}
      \rho_{S,\text{ss}}^\text{max} = \left[1+\frac{1}{2}\sqrt{\left(\frac{\kappa}{J}\right)^2+\frac{16}{C}} \right]^{-1}.
\end{equation}

There are two different limits to this maximum value that can be considered when approaching the optimum conditions of entanglement, namely (i) $C\gg \frac{J}{\kappa} \gg 1 $ or (ii) $\frac{J}{\kappa}\gg C \gg 1$. In the first case, we find that the maximum population for the entangled state is given by
\begin{equation}
\label{eq:rhoS_max_J}
    \rho_{S,\text{ss}}^\text{max}(C\gg J/\kappa) = \left(1+\frac{\kappa}{2J} \right)^{-1},
\end{equation}
while in the second case, 
\begin{equation}
\label{eq:rhoS_max_C}
    \rho_{S,\text{ss}}^\text{max}(J/\kappa \gg C) = \left(1+2/\sqrt{C} \right)^{-1}.
\end{equation}
Equations \eqref{eq:rhoS_max_J} and  \eqref{eq:rhoS_max_C} serve as guidelines that inform one about the maximum fidelity to $|S\rangle$ attainable (and thus the maximum values of stationary entanglement) when the limiting factor are given by the inter-emitter coupling rate, $J$, or the cavity cooperativity $C$, respectively.

\section{Dependence of the entanglement generation on the coupling rate between emitters}
\label{sec:SM_Validity}

\begin{figure}[t]
	\includegraphics[width=0.75\textwidth]{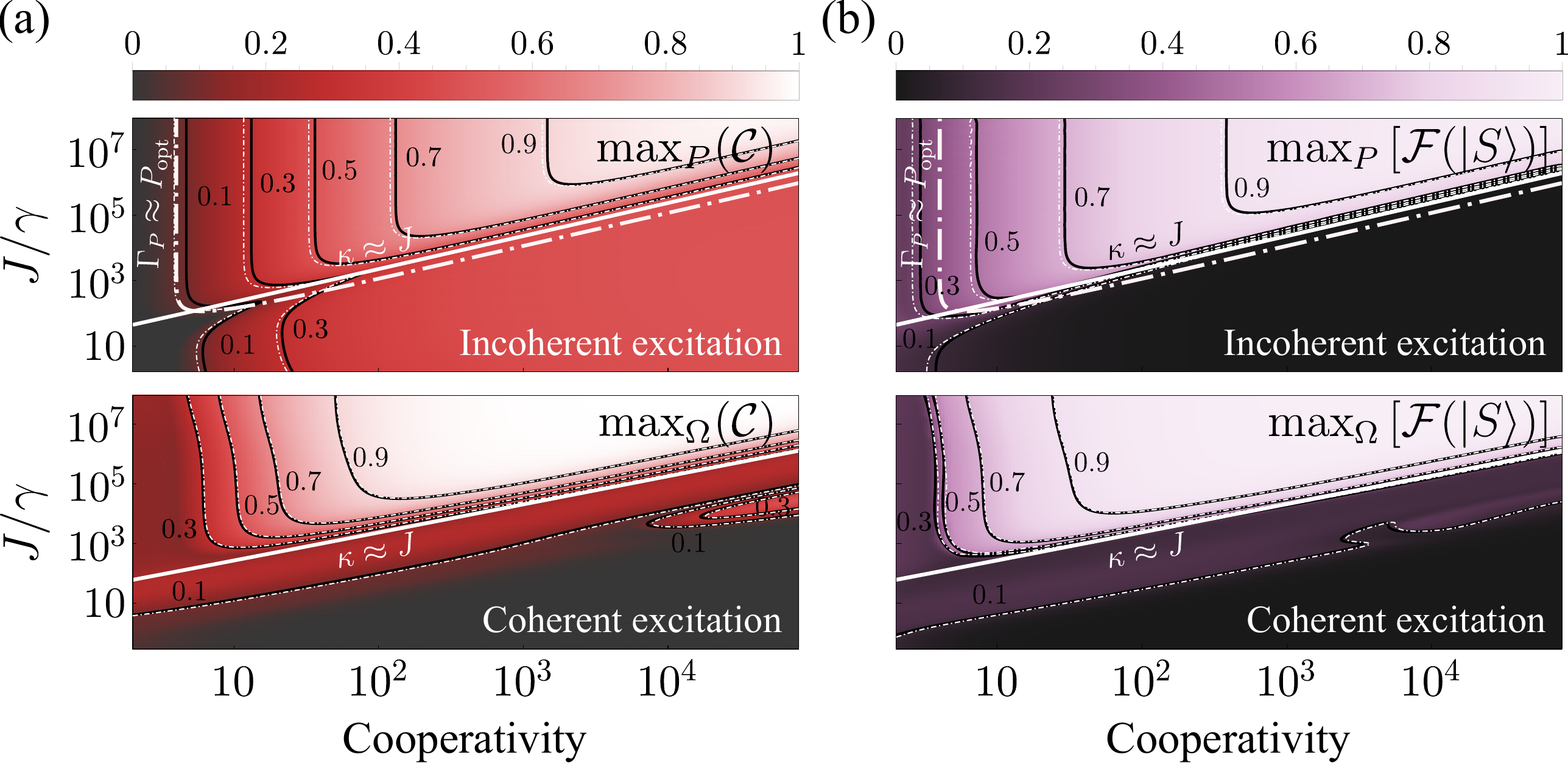}
	\caption{
 (a) Maximum achievable stationary concurrence and (b) fidelity with $|S\rangle$ versus cooperativity $C$ and dipole-dipole coupling $J$. 
 In each panel the driving intensity optimizes the generation of entanglement, that is, $P=P_{\text{opt}}$ and $\Omega=\Omega_{\text{opt}}$, respectively.
The black straight lines correspond to numerical computations, and grey-dashed lines are numerical predictions from the Bloch-Redfield master equation.
Parameters: $\delta=10^{-3}J$, $\Omega=\Omega_{\text{opt}}$, $P=P_{\text{opt}}$.
	}
	\label{fig:FigSM_Validity}
\end{figure}
The finite value of $J$ can be a limiting factor for the generation of entanglement, as we just saw in Eq.~\eqref{eq:rhoS_max_J}. Assuming optimal driving intensities, $P_{\text{opt}}$ and $\Omega_{\text{opt}}$, we extend here the numerical analysis presented in the main text to different values of the coupling rate between emitters, $J$.

In the simplest scenario, $J$ is set by dipole-dipole interactions, and therefore different value of $J$ correspond to different inter-emitter distances. With this situation in mind, we also vary $\gamma_{12}$ accordingly~\cite{ReitzCooperativeQuantum2022,FicekQuantumInterference2005,Vivas-VianaTwophotonResonance2021}. Nevertheless, we note that $\gamma_{12}$ is not a critical parameter (since we work in a regime in which it is overcome by the Purcell-enhanced decay rate $\Gamma_P\gg \gamma$) and that, in general, $J$ could be provided by mechanisms other than dipole-dipole interaction, e.g., mediated by photonic structures~\cite{goldstein1997,EvansPhotonmediatedInteractions2018,ChangColloquiumQuantum2018}.

In Fig.~\ref{fig:FigSM_Validity}, we show the stationary concurrence, $\text{max}_{P/\Omega}(\mathcal{C})$ and the fidelity to $|S\rangle$, $\text{max}_{P/\Omega}[\mathcal{F}(|S\rangle)]$, both maximized over the driving strength, plotted against $J$ and the cavity cooperativity $C$. 
In this plot, we see wide regions of values of $J$ and $C$ where high values of stationary entanglement are obtained, consistent with the regions where the symmetric state $|S\rangle$ is highly populated. We observe that these regions are well bounded by the condition $\kappa \ll J$ that allows for the frequency-resolved Purcell effect to take place.
In both schemes, we observe that interaction rates of at least $J\sim 10^3\gamma$ are necessary to achieve sizable values of the concurrence, with values as high as $\mathcal C\sim 0.7$ within reach for cooperativities in the range $C\sim 10-100$. 
These results underscore the generality of the reported mechanism and its potential application for obtaining maximum entanglement in different solid-state cavity QED platforms.

\section{Effect of additional decoherence channels}
\label{sec:SM_AdditionalDecoherence}

In the main text, the main sources of decoherence taken into account are local and collective spontaneous decay and cavity leakage, stemming from the interaction with a vacuum electromagnetic bath~\cite{FicekQuantumInterference2005,CarmichaelStatisticalMethods1999,Vivas-VianaDissipativeStabilization2023}. 
However, in order to assess whether our findings can be observed in realistic setups, 
additional decoherence channels must be considered~\cite{ShammahOpenQuantum2018}.
For instance, coupling to phonon baths or to lossy nanophotonic structures with specific geometries can yield additional spontaneous channels beyond the zero-phonon line~\cite{HaakhSqueezedLight2015, MlynekObservationDicke2014}, pure dephasing  (ubiquitous in semiconductor quantum dots or in molecular systems)~\cite{KrummheuerTheoryPure2002,delPinoQuantumTheory2015,ReitzMoleculephotonInteractions2020}; or collective pure dephasing~\cite{PrasannaVenkateshCooperativeEffects2018,WangDissipationDecoherence2015}.
To assess the robustness of our mechanism for generating high stationary entanglement, we  consider individually (for the case of two emitters $N=2$) the impact of additional decoherence in these three distinct scenarios: (i) extra spontaneous decay, (ii) local pure dephasing, and (iii) collective pure dephasing. 

These additional decoherence channels translate into three extra Lindblad terms in the master equation:
\begin{equation}
    \frac{d \hat \rho}{d t}=\mathcal{L}[\hat \rho]
+
\underbrace{\frac{\Gamma}{2}(\mathcal{D}[\hat \sigma_1]\hat \rho+\mathcal{D}[\hat \sigma_2]\hat \rho)}_{\substack{\text{Extra spontaneous} \\ \text{decay},\  \mathcal{L}_{\Gamma}[\hat \rho]}}
+
\underbrace{\frac{\gamma_\phi}{2}(\mathcal{D}[\hat \sigma_{z,1} ]\hat \rho+\mathcal{D}[\hat \sigma_{z,2} ]\hat \rho)}_{\text{Local pure dephasing},\ \mathcal{L}_{\gamma_\phi} [\hat \rho]}
+
\underbrace{\frac{\Gamma_\phi}{2}\mathcal{D}[\hat \sigma_{z,1}+\hat \sigma_{z,2} ]\hat \rho}_{\substack{\text{Collective pure} \\ \text{dephasing},\  \mathcal{L}_{\Gamma_\phi}[\hat \rho]}},
\end{equation}
where $\hat \sigma_{z,i}\equiv 2\hat \sigma_i^\dagger \hat \sigma_i-\mathbb{1}$ is the z-component of the Pauli matrices, and $\mathcal{L}[\hat \rho]$ denotes the original Liouvillian,
\begin{equation}
    \mathcal{L}[\hat \rho]\equiv -i [\hat H,\hat  \rho] 
+\frac{\kappa}{2}\mathcal{D}[\hat a]\hat \rho 
+ \sum_{i,j=1}^{N=2}\frac{\gamma_{ij}}{2}\mathcal{D}[\hat \sigma_i,\hat \sigma_j]\hat \rho +
\frac{P_i}{2}\mathcal{D}[\hat \sigma_i^\dagger ]\hat \rho.
\end{equation}
We will exhibit the effects of the additional decoherence channels independently, resulting in a master equation $d\hat \rho/dt=\mathcal{L}[\hat \rho]+\mathcal{L}_{\gamma_\text{d}}[\hat \rho]$, where $\gamma_\text{d}=\{\Gamma, \gamma_\phi, \Gamma_\phi\}$. Figure~\ref{fig:FigAppendix_Robustness}(a-c) depicts the concurrence versus the cooperativity and the additional decoherence rate for these three cases and both coherent and incoherent excitation schemes.
\subsection{General conditions}

\begin{figure}[b]
	\includegraphics[width=1.\textwidth]{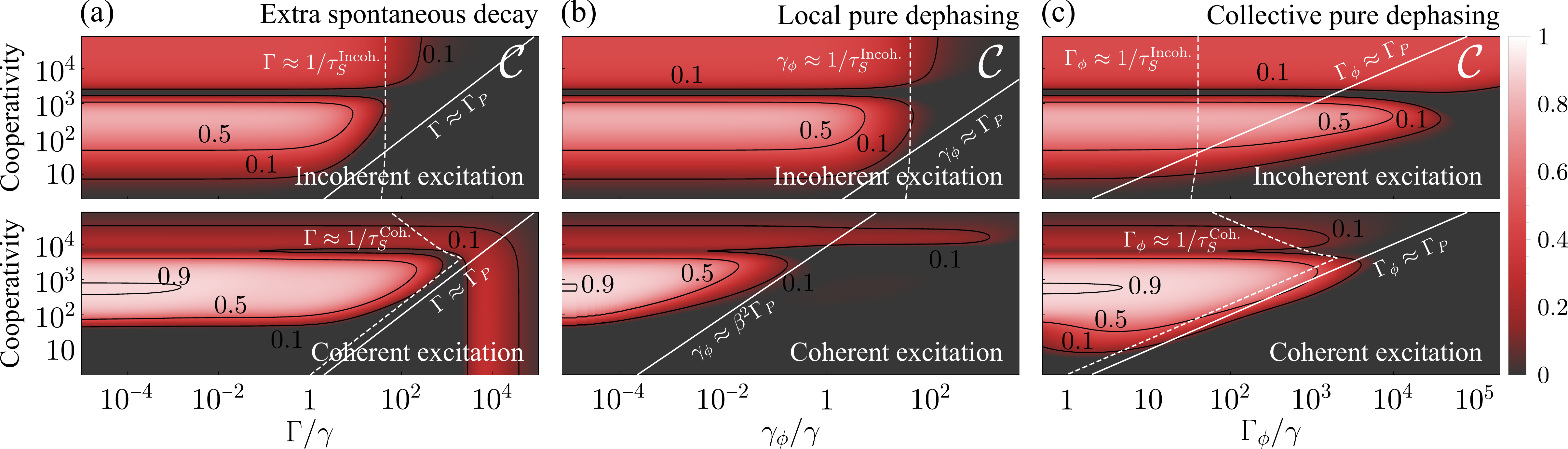}
	\caption{
 Robustness test of the mechanism of entanglement generation for both excitation schemes: incoherent (upper row) and coherent (lower row). The considered additional channels are: (a) extra spontaneous decay; (b) local pure dephasing; and (c) collective pure dephasing. Each panel depicts the concurrence as a function of the cavity cooperativity $C$ and the corresponding extra decoherence rate.
	Parameters: 
 $J=9.18\times 10^4\gamma,\ \gamma_{12}=0.999\gamma,\ \delta=10^{-2}J,\ \ g=10^{-1}\kappa,\ \Delta_a\approx -J,\ P=40\gamma,\ \Omega=10^4 \gamma$}
	\label{fig:FigAppendix_Robustness}
\end{figure}

The frequency-resolved Purcell effect is efficient when the Purcell rate $\Gamma_P$ is the dominant dissipative rate. 
Consequently, for the mechanism to be robust against additional decoherence, the Purcell rate should exceed the corresponding decoherence rates, i.e., $\Gamma_P\gg \gamma_{\text{d}}$.
This condition is illustrated in Fig.~\ref{fig:FigAppendix_Robustness}(a-c) by a solid-white line. This line bounds the region of entanglement in most cases, with some exceptions that we discuss below.
Additionally, for entanglement to survive in the long-time limit, the stabilization timescale $\tau_S$ must also be faster than the additional decoherence rates. This translates into the condition $\tau_S\ll \gamma_{\text{d}}^{-1}$, depicted in Fig.~\ref{fig:FigAppendix_Robustness}(a-c) by dashed white lines. 
In summary, we require:%
\begin{enumerate}[label=(\roman*)]
\item $\Gamma_P \gg \gamma_{\text{d}} $: The Purcell rate must dominate over all decoherence rates.
\item $\tau_S \ll \gamma_{\text{d}}^{-1}$: The stabilization timescale must be faster than the decoherence.
\end{enumerate}

There are two exceptions to these general observations: (i)  local pure dephasing with coherent drive; (ii)  collective pure dephasing with incoherent drive.

\subsection{Local pure dephasing with coherent excitation}

This first exception arises from the nature of the subradiant state, $|-\rangle$. As proven in Ref.~\cite{Vivas-VianaDissipativeStabilization2023}, the effective Lindblad term in the presence of coherent drive is given by a jump operator $\hat \xi_S=-|+\rangle \langle ee | +\beta/2|gg\rangle \langle -|$, resulting in
\begin{equation}
    \mathcal{L}_{\text{cav}}^{\text{eff}}=\Gamma_P\mathcal{D}[\hat \xi_S]\hat \rho \approx \Gamma_P \mathcal{D}[|+\rangle \langle ee| ]\hat \rho + \frac{(\beta^2/4) \Gamma_P}{2} \mathcal{D}[|gg\rangle \langle - |]\hat \rho.
\end{equation}
Local pure dephasing tends to mix the states $|+\rangle$ and $|-\rangle$ by transferring population from one to the other. In order for entanglement due to the population of $|+\rangle \approx |S\rangle$ to survive, the subradiant state $|-\rangle$ must be depleted faster than it is populated, which sets the condition $\beta^2 \Gamma_P \gg \gamma_\phi$. 
When this condition does not hold, the mechanism of entanglement generation breaks down since the population transfer between the super- and subradiant states occur with a rate $\propto \gamma_\phi$, destroying any possible entanglement.
Notably, the generation of entanglement exhibits greater robustness against local pure dephasing under incoherent excitation. In contrast to the coherent excitation scenario~\cite{Vivas-VianaDissipativeStabilization2023}, once the subradiant state acquires population due to the population transfer induced by this decoherence term, the incoherent excitation is able to deplete this state by repumping towards the doubly-excited state, leading it back into the symmetric state by means of the frequency-resolved Purcell effect.

If this condition does not hold, but the formation of the entangled state is faster than the decoherence timescale, $\tau_S\ll \gamma_\phi^{-1}$, we can still observe the formation of entanglement as a metastable state that survives for a time $ \gamma_\phi^{-1}$ (not shown).

\subsection{Collective pure dephasing and incoherent excitation}

The second exception is the specific case of collective pure dephasing and incoherent excitation, where the stationary entanglement is maintained even when the conditions (i-ii) do not hold. This type of decoherence essentially degrades the coherence between the ground state $|gg\rangle$ and the doubly-excited state $|ee\rangle$. Contrary to the coherent excitation scheme, in the case of incoherent excitation, population transfer among states occurs via incoherent one-photon processes, and does not rely on any type of coherence between $|gg\rangle$ and $|ee\rangle$ Therefore, the mechanism of entanglement generation via incoherent excitation proves to be highly robust against collective pure dephasing, maintaining entanglement even at very high values of this decoherence term ($\Gamma_\phi \sim 10^5 \gamma$). This type of dephasing only becomes detrimental when its rate becomes comparable to the coherent coupling, $J\approx \Gamma_\phi$, since then the relevant transitions are no longer resolved due to the dephasing-induced broadening.

\section{Entanglement in large emitter ensembles}
\label{sec:SM_AddingNEmitters}
In this Section, we provide further numerical evidence of the prospects of the proposed mechanism to generate entangled $W$ states in large emitter ensembles $N\gg 1$. To do this, we make use of the master equation  generalized to $N$ emitters,
\begin{equation}
    \frac{d \hat \rho}{dt}=-i[\hat H, \hat \rho]+\frac{\kappa}{2}\mathcal{D}[\hat a]\hat \rho +\sum_{i,j}^N \frac{\gamma_{ij}}{2}\mathcal{D}[\hat \sigma_i, \hat \sigma_j]\hat \rho + \sum_i^N \frac{P_i}{2} \mathcal{D}[\hat \sigma_i^\dagger]\hat \rho,
\end{equation}
where the Hamiltonian is now given by
\begin{equation}
    \hat H= J \hat S^+ \hat S^- + \Delta_a \hat a^\dagger \hat a +g(\hat a^\dagger \hat S^- +\hat  a \hat  S^+),
\end{equation}
having defined $S^-\equiv \sum_i^N \hat \sigma_i$. As it is considered in the case of $N=2$ emitters, $\gamma_{ii}=\gamma$, $P_i=P$, and $\gamma_{ij}=\gamma_{ji}$ (with $i\neq j$) for all $(i,j)=1,\ldots, N$, then we denote the collective dissipative coupling simply as $\gamma_{ij} \rightarrow \gamma_{\text{col}}$.

\begin{figure}[t]
	\includegraphics[width=1\textwidth]{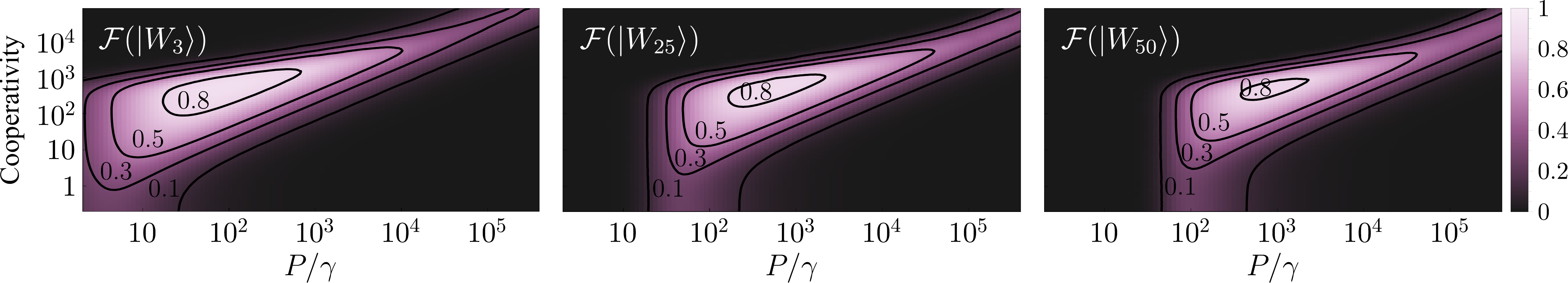}
	\caption{
 Fidelity to the $W$ state for $N=3, 25, 50$ emitters versus cavity cooperativity and incoherent pumping rate. 
Parameters: $J=10^5 \gamma,\ \gamma_{\text{col}}=0.999\gamma,\ g=10^{-1}\kappa,\ \Delta_a=J(2-N)$.
	}
	\label{fig:FigSM_2_Nqubits}
\end{figure}

In Fig.~\ref{fig:FigSM_2_Nqubits}, the fidelity $\mathcal{F}(|W_N\rangle)$ for $N=3,25, 50$ emitters is plotted versus cooperativity and the incoherent pumping rate $P$ when the cavity frequency is placed in resonance at $\Delta_a=J(N-2)$. These calculations make use of the PIQS library in QuTiP~\cite{ShammahOpenQuantum2018,johansson2012,johansson2013}).
Our numerical simulations confirm the stabilization of $W$ states with fidelities exceeding $0.8$ for numbers of emitters as high as $N=50$, which is the case shown in the main text.
The addition of extra emitters increases the minimum pumping rate required to generate entanglement. For instance, in the case of $N=3$, we observe that $P>\gamma$, while for $N=50$, a much higher pumping intensity is needed, $P\gtrsim 10^2\gamma$. This observation is consistent with the intuition that introducing more emitters increases the pumping intensity required to bring the system into a fully excited state $|ee\ldots e\rangle$, from where the relevant cavity-enhanced decay takes place.

\section{Entanglement based on post-selection}
\label{sec:SM_HeraldedMeasurements}
\begin{figure}[b]
	\includegraphics[width=1.\textwidth]{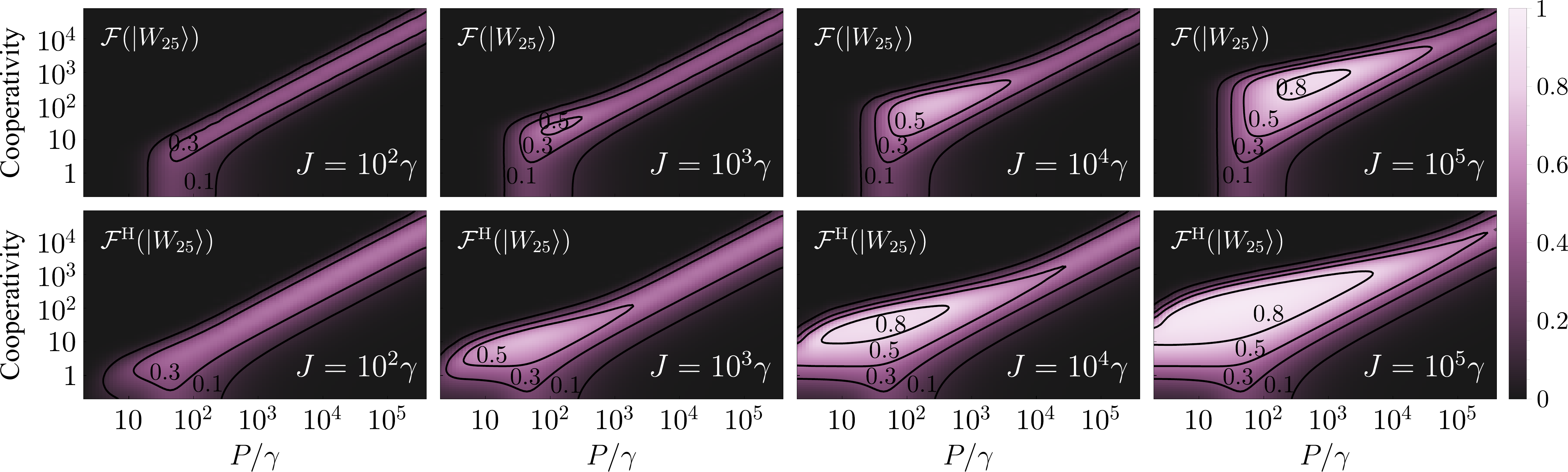}
	\caption{
 Comparison of fidelities between non- (upper row, $\mathcal{F}(|W_{25}\rangle)$) and post-selection (lower row, $\mathcal{F}^\text{H}(|W_{25}\rangle)$) measurement schemes for detecting $W$ states.
In both rows, the fidelity to the $W$ state for $N=25$ emitters is depicted versus cavity cooperativity and incoherent pumping rate for different coherent coupling rates are considered, from left to right, $J/\gamma=10^2, 10^3,10^4,10^5$.
Parameters: $\gamma_{\text{col}}=0.999\gamma,\ g=10^{-1}\kappa,\ \Delta_a=J(2-N)$.
	}
	\label{fig:FigSM_3_ComparisonNormalHeralded}
\end{figure}
Here we explore further the prospects of enhancing entanglement via post-selection---where the entangled state is heralded  by the detection of a photon emitted by the cavity---and how this allows to relax the condition of high coherent coupling between emitters.
For this, we perform calculations with a fixed number of emitters, $N=25$, and vary the coherent coupling rate from a relatively small value of $J=10^2\gamma$ to the case presented in the text of $J=10^5\gamma$. We then compare the fidelity to $|W\rangle$ obtained in the steady-state and heralded by the detection of a photon.

Figure~\ref{fig:FigSM_3_ComparisonNormalHeralded} illustrates the fidelity to the $W$ state versus cooperativity and incoherent pumping rate. The upper and lower rows represent the non-post-selected ($\mathcal{F}(|W_{25}\rangle)$) and post-selected ($\mathcal{F}^{H}(|W_{25}\rangle)$) detection schemes, respectively.
These results highlight the significant advantage of using post-selection measurement techniques to achieve higher fidelities to the $W$ state. For instance, in the case of $J=10^4$, the non-post-selected fidelity reaches a maximum achievable value of approximately $0.5$, while in the post-selected measurement, this quantity increases to around $0.8$
\section{Optical signatures of entanglement}
\label{sec:SM_ObservationalProperties}

\begin{figure}[t]
	\includegraphics[width=1.\textwidth]{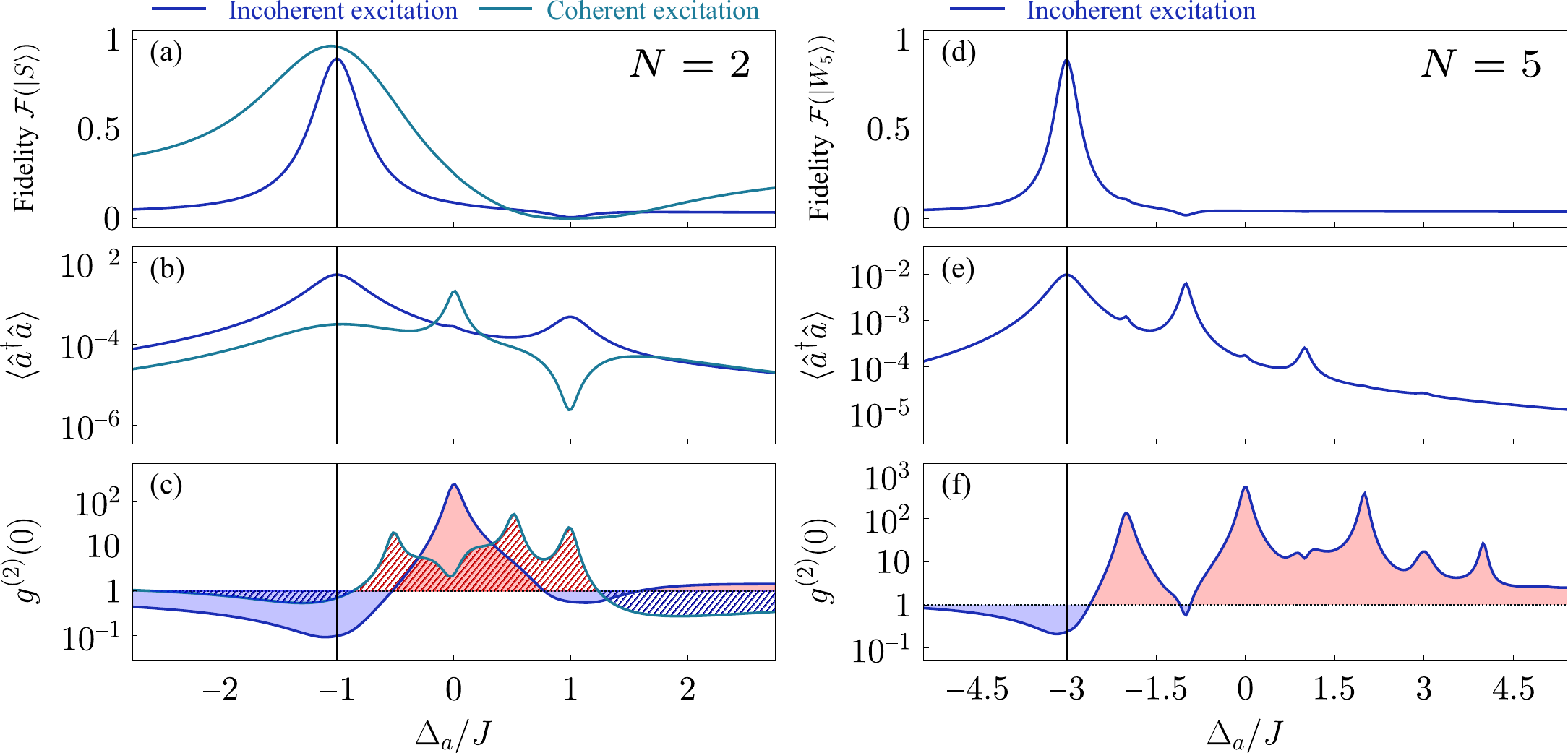}
	\caption{
 Correspondence between entanglement and the properties of the emitted light by the cavity. In both columns, corresponding to the case of $N=2$ (a-c) and $N=5$ (d-f) emitters, from top to bottom, the fidelity to the entangled state $|S\rangle$ (left) and $|W_5\rangle$, emission intensity $\langle \hat a^\dagger \hat a \rangle$, and second order correlations at zeroy delay $g^{2}(0)$, are computed versus cavity frequency $\Delta_a$.
 We note that in the case of $N=2$ emitters, both excitation schemes are illustrated. We use a solid filling for incoherent excitation and patterned filling for the coherent case.
 Parameters (Case $N=2$ qubits): (a-c)  $\ J=9.18\times 10^4\gamma$, $\gamma_{12}=0.999\gamma,$  $\delta=10^{-2}J$, $g=10^{-1}\kappa$, $C=500$, $P_{\text{opt}}=66.39 \gamma$, $\Omega_{\text{opt}}=8504.99 \gamma$; (d-f) $J=10^5 \gamma$, $\gamma_{\text{col}}=0.999\gamma$,  $g=10^{-1}\kappa$, $P_{\text{opt}}=132.3 \gamma$, $C=496.84$. 
	}
	\label{fig:FigSM_Observability}
\end{figure}

Here we show that the resonant activation of the frequency-resolved Purcell effect leading to entanglement correlates with features in the 
and classical and quantum observables of the light emitted by the system.
From top to bottom, Fig. \ref{fig:FigSM_Observability} depicts, for $N=2$ (left) and $N=5$ (right), the fidelity $\mathcal{F}(|W_N\rangle)$, the emission intensity $\langle \hat a^\dagger \hat a \rangle$, and the second-order correlations at zero delay $g^{(2)}(0)$.
We observe that a region of high fidelity---i.e., high degree of entanglement---correlates with a peak in the emission intensity, as well as an antibunching resonance in the photon statistics at frequencies $\Delta_a\approx -J, -3J$ for the case of $N=2,5$, respectively. In the case of incoherent pumping, the antibunching resonance remains even as the number of emitters is increased, in contrast with the general tendency of light emitted from an ensemble to thermalize. Optical signatures in the case coherent driving are discussed in more depth in Ref.~\cite{Vivas-VianaDissipativeStabilization2023}.

\end{document}